\documentclass[aps,prd,onecolumn,groupedaddress,nofootinbib,amssymb]{revtex4}

\usepackage[latin1]{inputenc}
\usepackage{graphicx}
\usepackage[english]{babel}

\begin{document}

\title{Quantum equation of motion for a particle in the field of primordial
fluctuations}

\author{S.N. Andrianov, V.V. Bochkarev, S.M. Kozyrev}

\affiliation{Scientific Center for Gravitational Wave Research
Dulkyn, Bauman 20, Kazan 420503, Russia}

\email {adrianovsn@mail.ru}

\begin{abstract}
Brane model of universe is considered for a particle. Conservation
laws inside the brane are obtained. Equation of motion is derived
for a particle using variation principle from these conservation
laws. This equation includes terms accounting the variation of
brane topology. Its solution is obtained at some approximations
and dispersion relation for a particle is derived.

 \vspace{4. mm}

Keywords: brane, quantum equation of motion, primordial
fluctuations
\end{abstract}

\maketitle

 \vspace{4. mm}

\section{Quantum equation of motion for a particle in the field of primordial
fluctuations} \label{s:intro}

The Klein-Gordon equation describing motion of a scalar particle
is known in quantum field theory that does not account the changes
of space metrics and changes of particles behavior connected with
it \cite{Ryder}. Variation of space-time metrics is described by
Einsteins equation. Wheeler de Witt equation occupies place of
Einsteins equation in quantum theory that is generalization for
the case of general relativity theory and is valid for arbitrary
Ryman space \cite{Wheeler}. The approach of Wheeler de Witt is
applied to brane theory of Universe in paper \cite{Darabi}.
However, the variation of brane topology is not accounted in this
paper. The variation of space topology is considered within
quantum theory in phenomenological way in paper \cite{Vilenkin}.
In the paper \cite{Koyama}, Wheeler de Witt equation is obtained
from priori taken action for inflating brane. In present paper, we
will derive starting from the symmetry properties of the brane the
equation of motion for massless particle in the framework of brane
model with the account of its topology variation in universal
space during primordial fluctuations considered in the framework
of classical theory in papers \cite{Wands}, \cite{Maartens}.

Lets consider our space as three-dimensional hyper-surface that is
the insertion in the space of higher dimension. Functional of
length (action at the movement of particle in universal space) can
be written as \cite{will1}

\begin{equation}
S=mc
\begin{tabular}{l}
T \\
$\int $ \\
0
\end{tabular}
ds,  \label{e1}
\end{equation}

where {\it m }is mass,{\it T } is current value of universal time,
{\it ds} is interval. Interval can be written as

\begin{equation}
{\it ds}=\sqrt{g_{ij}dx^idx^j-c^2dt^2},  \label{e2}
\end{equation}

where {\it i} and {\it j} numerate four coordinates ({\it
i,j}=0,1,2,3) of our space in the universal space, {\it t} is
universal time. Substituting expression (\ref{e2}) into formula
(\ref{e1}), we get

\begin{equation}
S=mc
\begin{tabular}{l}
T \\
$\int $ \\
0
\end{tabular}
\sqrt{g_{ij}dx^idx^j-c^2dt^2}.  \label{e3}
\end{equation}

Let's rewrite expression (\ref{e3}) in the form

\begin{equation}
S=\stackrel{}{
\begin{tabular}{l}
T \\
$\int $ \\
0
\end{tabular}
Ldt},  \label{e4}
\end{equation}

where

\begin{equation}
L=c\sqrt{g_{ij}m\stackrel{\cdot }{x}^im\stackrel{\cdot
}{x}^j-m^2c^2} \label{e5}
\end{equation}
is Lagrangian $\stackrel{\cdot }{\it x}^i=dx^i/dt$.

Let's suppose that lengths in universal space do not change at the
evolution of considered hypersurface, i.e. S$\left( T+\triangle
T\right) =S\left( T\right) .$ Then in correspondence with formula
(\ref{e4}) L=0 and we will have from expression (\ref{e5})

\begin{equation}
g_{ij}=p^ip^j=m^2c^2,  \label{e6}
\end{equation}

where p$^i=m\stackrel{\cdot }{x^i}\ $is component of particles
momentum at the movement on brane.

We can choose for two infinitely close events taking place in one
point of
our space such a reference system that {\it dx}$^1${\it \ =dx}$^2${\it \ =dx}%
$^3$ =0. Then, we can write accounting that {\it ds} = 0 the
following relation:

\begin{equation}
\sqrt{g_{00}}dx^0=cdt.  \label{e7}
\end{equation}

Relation (\ref{e7}) yields the expression for particles own time
that coincides with universal time

\begin{equation}
t=\frac 1c\int \sqrt{g_{00}}dx^0,  \label{e8}
\end{equation}

where integration is performed by coordinate of universal space
{\it x}$^0.$

In the case of universal space metrics $\left\{ +,-,-,-,-\right\}
$ where have set {\it cdt = x}$^4${\it \ }at the absence of
gravitational fields and accounting that interval between
infinitely close events is equal zero also in the arbitrary case,
we get common relation among the time counting by moving and still
clocks at the uniform straightforward motion of object
respectively us from point {\it A} to point {\it B}

\begin{equation}
dt=d\tau \sqrt{1-\frac{\nu ^2}{c^2}},  \label{e9}
\end{equation}

where $\nu ^2=\left( dx^1/d\tau \right) ^2$ + $\left( dx^2/d\tau
\right) ^2$ + $\left( dx^3/d\tau \right) ^2$, $d\tau =dx^0/c.$
Geometrical meaning of relation(\ref{e9}) is explained in fig. 1.

Expression (\ref{e6}) can be rewritten in the following form:

\begin{equation}
p_ip^j=m^2c^2.  \label{e10}
\end{equation}

Let's consider functional variation of relation (\ref{e10})
corresponding to
brane fluctuation when coordinares {\it x} transform into coordinates {\it x'%
} (fig. 2). Complete variation of momentum vector can be written
as the sum of functional variation $\delta p$ of vector {\it p} at
the comparison of {\it p'} with {\it p }in the same point at the
parallel transfer of vector {\it p} in universal space and
ordinary variation {\it dp}. Then, it can be written that

\begin{equation}
\triangle p=p^{\prime }\left( x^{\prime }\right) -p\left( x\right)
=p^{\prime }\left( x^{\prime }\right) -\stackrel{\sim }{p}\left(
x^{\prime }\right) +\stackrel{\sim }{p}\left( x^{\prime }\right)
-p\left( x\right) =\delta p+dp,  \label{e11}
\end{equation}

where

\begin{equation}
\delta p=p^{\prime }\left( x^{\prime }\right) -\ \stackrel{\sim
}{p}\left( x^{\prime }\right)  \label{e12}
\end{equation}

and

\begin{equation}
dp=\ \stackrel{\sim }{p}\left( x^{\prime }\right) -p\left(
x\right) , \label{e13}
\end{equation}

$\stackrel{\sim }{p}\left( x^{\prime }\right) \ $is momentum
vector at its parallel transfer in the universal space.

If trajectory of particle is geodetic one then

\begin{equation}
dp_i=\frac{\partial p_i}{\partial x^k}dx^k=0,  \label{e14}
\end{equation}

\begin{equation}
\delta p_i=\stackrel{\sim }{p}_k\Gamma _{il}^k\delta x^l,
\label{e15}
\end{equation}

where $\delta x^l=x^{\prime l}-x^l$ is variation in universal
space. Then, it can be written, omitting stroked index of momentum
vector,

\begin{equation}
p\left( x^{\prime }\right) =p\left( x\right) +\delta p.
\label{e16}
\end{equation}

At the transform {\it x }$\rightarrow ${\it \ x'}, relation
(\ref{e10}) is transforming accounting (\ref{e14}) to the
following form:

\begin{equation}
p_ip^i+p_i\delta p^i+\delta p_ip^i+\delta p_i\delta p^i=m^2c^2.
\label{e17}
\end{equation}

Let's pass in relation (\ref{e15}) to operators acting in Hilbert
space of wave functions $\psi \left( x\right) $. We represent for
this sake the components of vector {\it p} as

\begin{equation}
p_i=-i\hbar \frac \partial {\partial x^i},  \label{e18}
\end{equation}

and rewrite relation (\ref{e13}) as

\begin{equation}
\delta p_i=-\left\{ \Gamma _{i\alpha }^k\delta x^\alpha \right\}
_{;k}, \label{e19}
\end{equation}
where covariant derivative is performed in the point with stroked
indexes.

Let's consider the first term in the left side of equation
(\ref{e15}). For this purpose, we represent it in the form

\begin{equation}
p_ip^i=p_ig^{ij}p_j.  \label{e20}
\end{equation}

Using expression (\ref{e16}), we get

\begin{equation}
p_ip^i=-\hbar ^2\left( \frac{\partial g^{ij}}{\partial x^i}\frac
\partial {\partial x^j}+g^{ij}\frac{\partial ^2}{\partial
x^i\partial x^j}\right) . \label{e21}
\end{equation}

Let's use well known relation

\begin{equation}
\frac{\partial g^{ij}}{\partial x^k}=-\Gamma _{mk}^ig^{mj}-\Gamma
_{mk}^jg^{im}.  \label{e22}
\end{equation}

Then

\begin{equation}
p_ip^i=-\hbar ^2\left( g^{ij}\frac{\partial ^2}{\partial x^i\partial x^j}%
-g^{mj}\Gamma _{mk}^i\frac \partial {\partial x^j}-g^{im}\Gamma
_{mk}^j\frac
\partial {\partial x^i}\right) .  \label{e23}
\end{equation}
Changing indexes of summation, we get

\begin{equation}
p_ip^i=-\hbar ^2g^{ij}\left( \frac{\partial ^2}{\partial x^i\partial x^j}%
-\Gamma _{ik}^k\frac \partial {\partial x^j}-\Gamma _{ij}^k\frac
\partial {\partial x^k}\right) .  \label{e24}
\end{equation}
Let's consider second term in the left side of equation
(\ref{e15}), rewriting it in the form

\begin{equation}
p_i\delta p^i=p_ig^{ij}\delta p_j.  \label{e25}
\end{equation}

Using formula (\ref{e16}), we get

\begin{equation}
p_i\delta p^i=g^{ij}p_i\delta p_j+i\hbar \left( g^{ij}\Gamma
_{im}^m+g^{im}\Gamma _{im}^j\right) \delta p_j.  \label{e26}
\end{equation}

Let's write in its direct form the covariant derivative in the expression (%
\ref{e16}):

\begin{equation}
\delta p_j=-i\hbar \left( \Gamma _{jk}^i+\frac{\partial \Gamma _{jl}^k}{%
\partial x^k}\delta x^l-\Gamma _{jk}^n\Gamma _{nl}^k\delta x^l+\Gamma
_{nk}^k\Gamma _{jl}^n\delta x^l\right) ,  \label{e27}
\end{equation}

where stroked index of the derivative on is omitted. Substituting
expression (\ref{e27}) into formula (\ref{e26}), we get

\begin{eqnarray}
&&
\begin{tabular}{l}
$p_i\delta p^i=\hbar ^2g^{ij}(\frac{\partial \Gamma _{jk}^k}{\partial x^i}-%
\frac{\partial \Gamma _{ij}^k}{\partial x^k}+\Gamma _{in}^k\Gamma
_{jk}^n-2\Gamma _{ij}^n\Gamma _{nk}^k-\Gamma _{im}^m\Gamma _{jk}^k+\frac{%
\partial ^2\Gamma _{jl}^k}{\partial x^i\partial x^k}\delta x^l-$ \\
\\
$-\Gamma _{jk}^n\frac{\partial \Gamma _{nl}^k}{\partial x^i}\delta
x^l-\Gamma _{nl}^k\frac{\partial \Gamma _{jk}^n}{\partial
x^i}\delta x^l+\Gamma _{nk}^k\frac{\partial \Gamma
_{jl}^n}{\partial x^i}\delta x^l+\Gamma _{jl}^n\frac{\Gamma
_{nk}^k}{\partial x^i}\delta x^l-\Gamma
_{im}^m\frac{\partial \Gamma _{jl}^k}{\partial x^k}\delta x^l-$ \\
\\
$-\Gamma _{ij}^m\frac{\partial \Gamma _{lm}^k}{\partial x^k}\delta
x^l+\Gamma _{im}^m\Gamma _{jk}^n\Gamma _{nl}^k\delta x^l-\Gamma
_{im}^m\Gamma _{nk}^k\Gamma _{jl}^n\delta x^l+\Gamma _{ij}^m\Gamma
_{mk}^n\Gamma _{\ln }^k\delta x^l$ \\
\\
$-\Gamma _{ij}^m\Gamma _{nk}^k\Gamma _{lm}^n\delta x^l).$%
\end{tabular}
\label{e28}
\end{eqnarray}

Third term in the left side of equation (\ref{e15})

\begin{equation}
\delta p_ip^i=\delta p_ig^{ij}p_j,  \label{e29}
\end{equation}

has according to expressions (\ref{e16}), (\ref{e27}) the
following form:

\begin{equation}
\delta p_ip^i=-\hbar ^2g^{ij}\left( \Gamma _{ik}^k+\frac{\partial
\Gamma _{ij}^k}{\partial x^k}\delta x^l-\Gamma _{ik}^n\Gamma
_{nl}^k\delta x^l+\Gamma _{nk}^k\Gamma _{il}^n\delta x^l\right)
\frac \partial {\partial x^j}.  \label{e30}
\end{equation}

The last term in the left side of equation (\ref{e15})

\begin{equation}
\delta p_i\delta p^i=\delta p_ig^{ij}\delta p_j,  \label{e31}
\end{equation}

can be written with the account of expression (\ref{e27}) in the
form

\begin{eqnarray}
\begin{tabular}{l}
$\delta p_i\delta p^i=-\hbar ^2g^{ij}(\Gamma _{jk}^k\Gamma
_{jn}^n+\Gamma
_{jk}^k\frac{\partial \Gamma _{jl}^n}{\partial x^n}\delta x^l+\Gamma _{jn}^k%
\frac{\partial \Gamma _{il}^k}{\partial x^k}\delta x^l-$ \\
\\
$-\Gamma _{jk}^k\Gamma _{jn}^m\Gamma _{ml}^n\delta x^l+\Gamma
_{ik}^k\Gamma _{jl}^m\Gamma _{mn}^n\delta x^l-\Gamma _{ik}^m\Gamma
_{ml}^k\Gamma
_{jn}^n\delta x^l+$ \\
\\
$+\Gamma _{jl}^m\Gamma _{mk}^k\Gamma _{jn}^n\delta
x^l+\frac{\partial \Gamma
_{il}^k}{\partial x^k}\frac{\partial \Gamma _{jl^{\prime }}^n}{\partial x^n}%
\delta x^l\delta x^{l^{\prime }}-\frac{\partial \Gamma _{il}^k}{\partial x^k}%
\Gamma _{jn}^m\Gamma _{ml^{\prime }}^n\delta x^l\delta x^{l^{\prime }}+$ \\
\\
+$\frac{\partial \Gamma _{il}^k}{\partial x^k}\Gamma _{jl^{\prime
}}^m\Gamma _{mn}^n\delta x^l\delta x^{l^{\prime }}-\frac{\partial
\Gamma _{jl^{\prime }}^n}{\partial x^n}\Gamma _{ik}^m\Gamma
_{ml}^k\delta x^l\delta x^{l^{\prime }}+\frac{\partial \Gamma
_{jl^{\prime }}^n}{\partial x^k}\Gamma
_{il}^m\Gamma _{mk}^k\delta x^l\delta x^{l^{\prime }}+$ \\
\\
$\Gamma _{ik}^m\Gamma _{ml}^k\Gamma _{jn}^{m^{\prime }}\Gamma
_{m^{\prime }l^{\prime }}^n\delta x^l\delta x^{l^{\prime }}-\Gamma
_{ik}^m\Gamma _{ml}^k\Gamma _{jl^{\prime }}^{m^{\prime }}\Gamma
_{m^{\prime }n}^n\delta
x^l\delta x^{l^{\prime }}-$ \\
\\
$-\Gamma _{il}^m\Gamma _{mk}^k\Gamma _{jn}^{m^{\prime }}\Gamma
_{m^{\prime }l^{\prime }}^n\delta x^l\delta x^{l^{\prime }}+\Gamma
_{il}^m\Gamma _{mk}^k\Gamma _{jl^{\prime }}^{m^{\prime }}\Gamma
_{m^{\prime }n}^n\delta
x^l\delta x^{l^{\prime }}).$%
\end{tabular}
\label{e32}
\end{eqnarray}

Substituting expressions (\ref{e24}), (\ref{e28}), (\ref{e29}) and (\ref{e32}%
) into equation (\ref{e15}), we get equation for the wave function
$\psi $

\begin{eqnarray}
\begin{tabular}{l}
$g^{ij}[\left( \frac{\partial ^2}{\partial x^i\partial x^j}-\Gamma
_{ij}^k\frac \partial {\partial x^k}-\Gamma _{ij}^m\Gamma
_{mk}^k\right) \psi +\left( \frac{\partial \Gamma
_{jk}^k}{\partial x^i}-\frac{\partial \Gamma _{il}^k}{\partial
x^k}+\Gamma _{in}^k\Gamma _{jk}^n-\Gamma
_{ij}^n\Gamma _{nk}^k\right) \psi -$ \\
\\
$-\delta (\Gamma _{jk}^n\frac{\partial \Gamma _{nl}^k}{\partial
x^i}+\Gamma
_{nl}^k\frac{\partial \Gamma _{jk}^n}{\partial x^i}-\Gamma _{nk}^k\frac{%
\partial \Gamma _{jl}^n}{\partial x^i}-\Gamma _{jl}^n\frac{\partial \Gamma
_{nk}^k}{\partial x^i}+\Gamma _{im}^m\frac{\partial \Gamma
_{jl}^k}{\partial
x^k}+\Gamma _{ij}^m\frac{\partial \Gamma _{lm}^k}{\partial x^k}-\frac{%
\partial ^2}{\partial x^i\partial x^k}-$ \\
\\
$-\Gamma _{im}^m\Gamma _{jk}^n\Gamma _{nl}^k+\Gamma _{im}^m\Gamma
_{nk}^k\Gamma _{jl}^n-\Gamma _{ij}^m\Gamma _{mk}^n\Gamma
_{nl}^k+\Gamma _{ij}^m\Gamma _{nk}^k\Gamma _{lm}^n+\Gamma
_{jk}^k\Gamma _{jn}^m\Gamma
_{ml}^n-$ \\
\\
$-\Gamma _{jk}^k\Gamma _{jl}^m\Gamma _{mn}^n-\Gamma _{ik}^m\Gamma
_{ml}^k\Gamma _{jn}^n+\Gamma _{jl}^m\Gamma _{mk}^k\Gamma
_{jn}^n)\Psi +\delta x^l\left( \frac{\partial \Gamma
_{il}^k}{\partial x^k}-\Gamma _{ik}^n\Gamma _{nl}^k+\Gamma
_{nk}^k\Gamma _{il}^n\right) \frac{\partial
\Psi }{\partial x^i}+$ \\
\\
$\delta x^l\delta x^{l^{\prime }}(\frac{\partial \Gamma _{il}^k}{\partial x^k%
}\frac{\partial \Gamma _{jl^{\prime }}^n}{\partial
x^n}-\frac{\partial
\Gamma _{il}^k}{\partial x^k}\Gamma _{jn}^m\Gamma _{ml^{\prime }}^n+\frac{%
\partial \Gamma _{il}^k}{\partial x^k}\Gamma _{jl^{\prime }}^m\Gamma _{mn}^n-%
\frac{\partial \Gamma _{il}^k}{\partial x^n}\Gamma _{ik}^n\Gamma _{ml}^k+%
\frac{\partial \Gamma _{il^{\prime }}^n}{\partial x^n}\Gamma
_{il}^m\Gamma
_{mk}^k+$ \\
\\
$+\Gamma _{ik}^m\Gamma _{ml}^k\Gamma _{jn}^{m^{\prime }}\Gamma
_{m^{\prime }l^{\prime }}^n-\Gamma _{ik}^m\Gamma _{ml}^k\Gamma
_{jl^{\prime }}^{m^{\prime }}\Gamma _{m^{\prime }n}^n-\Gamma
_{il}^m\Gamma _{mk}^k\Gamma
_{jn}^{m^{\prime }}\Gamma _{l^{\prime }m^{\prime }}^n+$ \\
\\
+$\Gamma _{il}^m\Gamma _{mk}^k\Gamma _{jl^{\prime }}^{m^{\prime
}}\Gamma
_{m^{\prime }n}^n)]\Psi +m^2c^2=0$%
\end{tabular}
\label{e33}
\end{eqnarray}

Let's consider covariant derivative of the second order for the
wave function

\begin{equation}
\left\{ \left\{ \psi ^n\right\} _{;i}\right\} _{;j}=\left\{
\frac{\partial \psi ^n}{\partial x^i}+\Gamma _{ik}^n\left(
x^{\prime }\right) \psi ^k\left( x^{\prime }\right) \right\}
_{;j}=\frac{\partial ^2\psi ^n}{\partial x^i\partial x^j}-\Gamma
_{ij}^k\frac{\partial \psi ^n}{\partial x^k}-\Gamma _{ij}^m\Gamma
_{mk}^n\psi ^k\left( x^{\prime }\right) .  \label{e34}
\end{equation}

If transformed wave function is still self function of energy
operator, i.e. if with account of norm the relation

\begin{equation}
\psi ^n\left( x\right) =\left( \Gamma _{ik}^k\right) ^{-1}\Gamma
_{ik}^n\psi ^k\left( x^{\prime }\right) ,  \label{e35}
\end{equation}

is true, the first term in the equation (\ref{e33}) can be written
in the form of covariant D'Alambertian

\begin{equation}
g^{ij}\left( \frac{\partial ^2}{\partial x^i\partial x^j}-\Gamma
_{ij}^k\frac \partial {\partial x^k}-\Gamma _{ij}^m\Gamma
_{mk}^n\right) \psi =g^{ij}D_iD_j\psi .  \label{e36}
\end{equation}

It is evident that the expression in brackets in the second term
in the left side of equation (\ref{e33}) is just Richey's tensor
and we can write

\begin{equation}
g^{ij}\left( \frac{\partial \Gamma _{jk}^k}{\partial
x^i}-\frac{\partial \Gamma _{ij}^k}{\partial x^k}+\Gamma
_{in}^k\Gamma _{jk}^n-\Gamma _{ij}^n\Gamma _{nk}^k\right) \psi
=-g^{ij}R_{ij}\psi .  \label{e37}
\end{equation}

We assume that $\delta x^l=0\ $for coordinates on brane and
$\delta x^l\neq 0\ $for coordinates along the additional
dimension. Then, assuming that coefficients of affine connection
on brane do not depend on the coordinate of additional dimension,
it could be written

\begin{equation}
g^{ij}\left( \Gamma _{jk}^n\frac{\partial \Gamma _{nl}^k}{\partial x^i}%
+\Gamma _{nl}^k\frac{\partial \Gamma _{jk}^n}{\partial x^i}-\Gamma _{nk}^k%
\frac{\partial \Gamma _{jl}^n}{\partial x^i}-\Gamma
_{jl}^n\frac{\partial \Gamma _{nk}^k}{\partial x^i}-\frac{\partial
^2\Gamma _{jl}^k}{\partial x^i\partial x^k}\right)
=-g^{ij}\frac{\partial R_{jl}}{\partial x^i}, \label{e38}
\end{equation}

\begin{equation}
g^{ij}\left( \Gamma _{ij}^m\frac{\partial \Gamma _{lm}^k}{\partial x^k}%
-\Gamma _{ij}^m\Gamma _{mk}^n\Gamma _{l\ n}^k+\Gamma _{ij}^m\Gamma
_{nk}^k\Gamma _{lm}^n\right) =g^{ij}\Gamma _{ij}^mR_{ml},
\label{e39}
\end{equation}

\begin{equation}
g^{ij}\left( \Gamma _{im}^m\frac{\partial \Gamma _{jl}^k}{\partial x^k}%
-\Gamma _{im}^m\Gamma _{jk}^n\Gamma _{nl}^k+\Gamma _{im}^m\Gamma
_{nk}^k\Gamma _{jl}^n\right) =g^{ij}\Gamma _{im}^mR_{jl},
\label{e40}
\end{equation}

\begin{equation}
g^{ij}\left( \frac{\partial \Gamma _{il}^k}{\partial x^k}-\Gamma
_{ik}^n\Gamma _{nl}^k+\Gamma _{nk}^k\Gamma _{il}^n\right)
=g^{ij}R_{il}. \label{e41}
\end{equation}

Relations (\ref{e38}),(\ref{e39}) yield

\begin{equation}
g^{ij}\frac{\partial R_{jl}}{\partial x^i}-g^{ij}\Gamma
_{ij}^mR_{ml}=g^{ij}D_iR_{jl}.  \label{e42}
\end{equation}

Using the same approach that at the derivation of relation
(\ref{e36}), we get from formulas (\ref{e40}),(\ref{e41})

\begin{equation}
g^{ij}R_{il}\frac{\partial \psi _n}{\partial x^i}-g^{ij}\Gamma
_{im}^mR_{il}\psi _n=g^{ij}R_{jl}D_i\psi _n.  \label{e43}
\end{equation}

Unifying expressions (\ref{e42}),(\ref{e43}), we get

\begin{equation}
g^{ij}\left( D_iR_{il}\right) \psi _n-g^{ij}R_{il}D_i\psi
_n=g^{ij}D_i\left( R_{jl}\psi \right) .  \label{e44}
\end{equation}

Last term in the left side of equation (\ref{e33}) can be written
as

\begin{equation}
g^{ij}\left( \frac{\partial \Gamma _{il}^k}{\partial x^k}-\Gamma
_{ik}^m\Gamma _{ml}^k+\Gamma _{mk}^k\Gamma _{il}^m\right) \left( \frac{%
\partial \Gamma _{jl^{\prime }}^n}{\partial x^n}-\Gamma _{jn}^{m^{\prime
}}\Gamma _{m^{\prime }l^{\prime }}^n+\Gamma _{m^{\prime
}n}^n\Gamma _{jl^{\prime }}^{m^{\prime }}\right) \delta x^l\delta
x^{l^{\prime }}\psi =g^{ij}R_{il}R_{il^{\prime }}\delta x^l\delta
x^{l^{\prime }}\psi . \label{e45}
\end{equation}

Thus, equation (\ref{e33}) can be rewritten in the following form:

\begin{equation}
g^{ij}D_iD_j\psi _n-g^{ij}R_{il}D_i\psi _n+g^{ij}D_i\left(
R_{jl}\psi _n\right) \delta x^l+g^{ij}R_{il}R_{il^{\prime }}\delta
x^l\delta x^{l^{\prime }}\psi _n+m^2c^2=0..  \label{e46}
\end{equation}

Let's consider small region of space-time where we can suppose
gravitation field to be constant and homogeneous. We rewrite
equation (\ref{e46}) in locally-geodesic coordinate system for
mass-less particle in the following form:

\begin{equation}
\left\{ \frac{\partial ^2}{\partial x^2}+\gamma \frac \partial
{\partial x}+\alpha \right\} \psi =\frac 1{c^2}\frac{\partial
^2\psi }{\partial t^2}, \label{e47}
\end{equation}

limiting ourselves by one spatial dimension, assuming the absence
of affine connection in time and introducing the notations

\begin{equation}
\gamma =R_{x\theta }\delta x^\theta  \label{e48}
\end{equation}

and

\begin{equation}
\alpha =R-\left( \frac \partial {\partial x}R_{x\theta }\right)
\delta x^\theta -R_{x\theta }R_{x\theta ^{\prime }}\delta x^\theta
\delta x^{\theta ^{\prime }}.  \label{e49}
\end{equation}

Let's look for solution in the form

\begin{equation}
\psi =e^{i\left( kx-\omega t\right) }.  \label{e50}
\end{equation}

Then we get from (\ref{e47}) the characteristic equation

\begin{equation}
k^2-i\gamma k-\left( \alpha +\frac{\omega ^2}{c^2}\right) =0.
\label{e51}
\end{equation}

Its physically reasonable solution is

\begin{equation}
k=\sqrt{\frac{\omega ^2}{c^2}+\alpha -\frac{\gamma ^2}4}+i\frac
\gamma 2. \label{e52}
\end{equation}

Apparently it is valid when a does not depend on coordinate and time. When $%
\alpha =\gamma =0,$ we have the usual dispersion relation $\omega
_0=ck.$

When $\left( \frac \omega c\right) \gg \alpha -\left( \frac \gamma
2\right) ^2$we can approximately get

\begin{equation}
k=\frac \omega c\left( 1+\frac{c^2}{\omega ^2}\left( \alpha -\frac{\gamma ^2}%
4\right) \right) +i\frac \gamma 2  \label{e53}
\end{equation}

and for phase velocity of non-massive particles

\begin{equation}
\frac{dk}{d\omega }=c\left( 1+\frac{c^2}{\omega ^2}\left( \alpha -\frac{%
\gamma ^2}4\right) \right) .  \label{e54}
\end{equation}

Formula (\ref{e54}) shows that effective refractive index related
with space curvature is equal

\begin{equation}
n_{eff}=1-\frac{c^2}{\omega ^2}\left( \alpha -\frac{\gamma
^2}4\right) . \label{e55}
\end{equation}

Thus, the phase velocity of mass-less particles in universal space
can exceed the phase velocity of light in plane Deckard space
because of drift of particles at the expansion of Universe. Other
consequences of space
curvature are the following two facts that realize when expression (\ref{e55}%
) is valid:

- it is impossible to create particle with kinetic energy less
than $\hbar \sqrt{\omega ^2+\alpha c^2}$ (fig. 3);

- space curvature leads to the frequency shift according the
formula $\omega =\sqrt{\omega _0^2+\alpha c^2}$ that gives the
possibility for verification of developed model when curvature
varies at the influence of gravitational waves and primordial
fluctuations.

Note that it could also represent the more complexity solution of
equation (47) in the form:

\begin{equation}
\psi =e^{-\frac 12x\left( \gamma -\sqrt{\gamma ^2-4\alpha }\right)
}\left( \alpha +\beta \ e^{x\left( \gamma -\sqrt{\gamma ^2-4\alpha
}\right) }\right) +A\ e^{ct\sqrt{\alpha }}+B\ e^{-ct\sqrt{\alpha
}}.  \label{e56}
\end{equation}

where A, B and $\beta $ are an arbitrary constants of integration.


\begin{figure}[ht]
\includegraphics[scale=0.7]{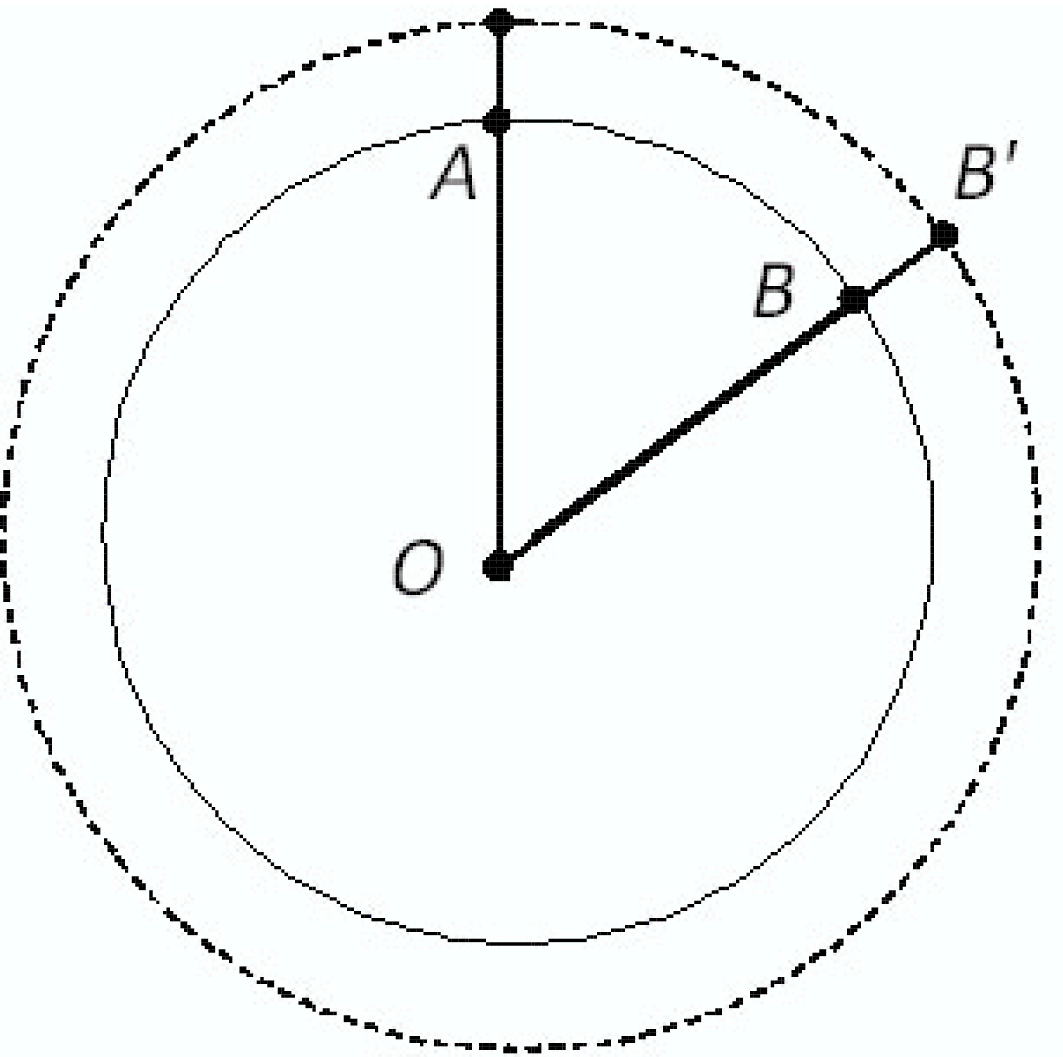}
\caption{ Brane model of Universe. Motion of particle from point A
to point B(B') in broadening Universe ( AB=$\nu \ d\tau ,\
AB^{\prime }=c\ d\tau ,\ BB^{\prime }=c\ dt$ ).} \label{fig1}
\end{figure}

\begin{figure}[ht]
\includegraphics[scale=0.7]{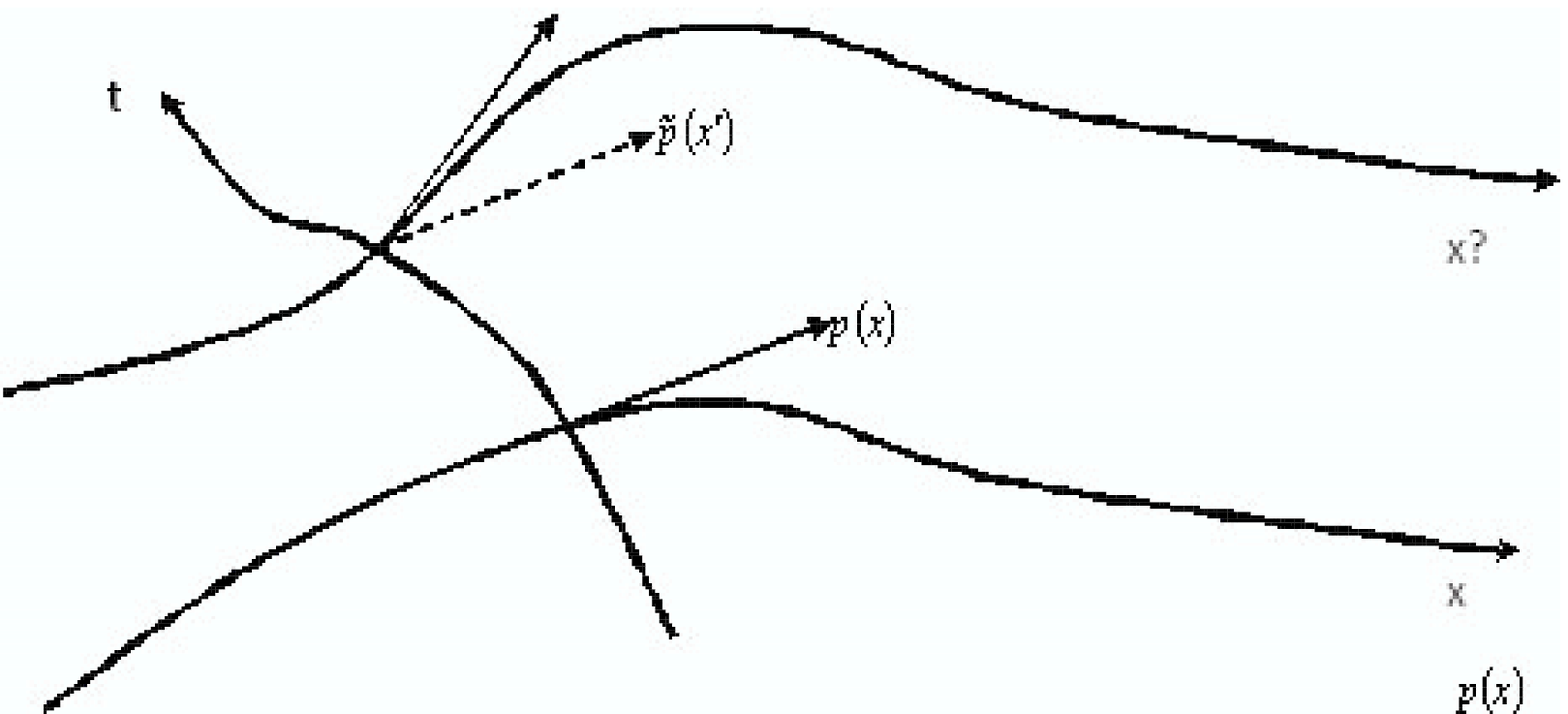}
\caption{Coordinate transform at the excitation of membrane.}
\label{fig:og}
\end{figure}

\begin{figure}[ht]
\includegraphics[scale=0.7]{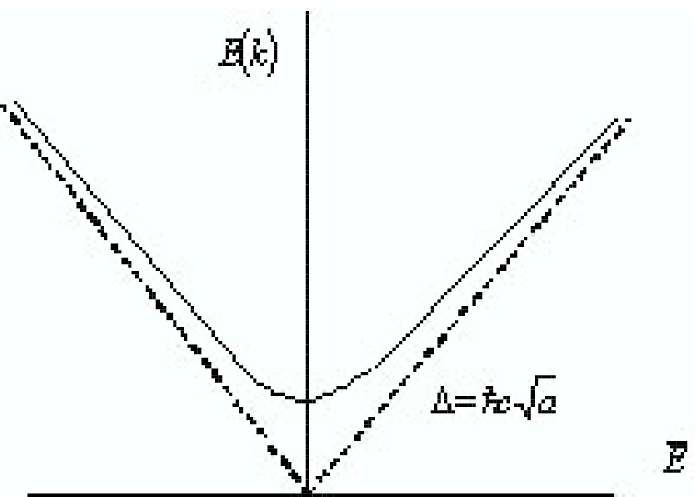}
\caption{Energy of moving particle E(k)=ck. } \label{fig:og}
\end{figure}
\end{document}